\documentclass[a4paper,11pt]{article}
\usepackage{jheppub}
\usepackage{bbm}

\newtheorem{theorem}{Theorem}
\newtheorem{observation}[theorem]{Observation}

\newcommand{\one}{\mathbbm{1}}
\newcommand{\Tr}[1]{\mathrm{Tr}\left[ {#1} \right]}

\newcommand{\qed}{\ensuremath{\hfill \Box}}
\newcommand{\ie}{i.e.\ }

\newcommand{\ket}[1]{|#1\rangle}

\newcommand{\ketbra}[2]{| #1 \rangle \langle #2 |}

\begin{document}

\title{Equilibration in low-dimensional quantum matrix models}  

\author[a]{R. H{\"u}bener,}
\author[b]{Y. Sekino,}
\author[a]{and J. Eisert}

\affiliation[a]{Dahlem Center for Complex Quantum Systems, Freie
Universit\"at Berlin, Arnimallee 14, 14195 Berlin, Germany}
\affiliation[b]{KEK Theory Center,  High Energy Accelerator Research
Organisation, 1-1 Oho, Tsukuba 305-0801, Japan}

\abstract{Matrix models play an important role in studies of quantum gravity, being candidates for a formulation of M-theory, but are notoriously difficult to solve. In this work, we present a fresh approach by introducing a novel exact model, provably equivalent with a low-dimensional bosonic matrix model, which is on its own a well-known, unsolved model of quantum chaos. In our equivalent reformulation local structure becomes apparent, facilitating analytical and precise numerical study. We derive a substantial part of the low energy spectrum, find a conserved charge, and are able to derive numerically the Regge trajectories. To exemplify the usefulness of the approach, we address questions of equilibration starting from a non-equilibrium situation, building upon an intuition from quantum information. We finally discuss possible generalisations of the approach.}


\maketitle


\section{Introduction}

Matrix quantum mechanics has received significant attention in recent years, mainly for its suspected connection with quantum gravity. Most prominently, Banks, Fischler, Shenker and Susskind (BFSS)~\cite{BFSS96} have proposed that supersymmetric matrix quantum mechanics, called matrix theory, gives a formulation of M-theory. This is an exciting proposal about the fundamental degrees of freedom in nature, which may allow us to address questions that are out of reach of semi-classical gravity. Questions of particular interest are on the quantum properties of black holes, 
such as their microscopic constituents, thermalisation, and the process of evaporation~\cite{HaPr07,Susskind}.

The bosonic degrees of freedom of BFSS matrix theory are $N\times N$ Hermitian matrices $\mathfrak{X}^{(i)}$, where $i=1,\ldots, d$ is the spatial index, and the dimension of space for the BFSS theory is $d=9$. There is no mass term, and the interaction is given by a quartic term. Unsurprisingly, this is an intricate theory and it has not been solved to date. To start with, every local degree of freedom directly couples to any other degree of freedom. What is more, the strong quartic interactions render na\"ive approximation schemes such as perturbation theory hopeless.

The primary focus of this article is on the $d=2$, $N=2$ model, as a
first paradigmatic step. Even though the original proposal of BFSS
concerns a large $N$ limit, the understanding of $N=2$ models is 
important since one can compute the scattering of two D0-branes
with $N=2$. 
This will allow us to probe the short-distance physics at the
$11$-dimensional Planck scale~\cite{DKPS97}. When the D0-branes are far
apart, a perturbative method based on the Born-Oppenheimer approximation
is valid~\cite{OkYo99, Taylor}, but as the D0-branes approach each
other, they enter the strong coupling regime where open strings between
them are excited. A qualitative discussion of D0-brane scattering has
been given in~\cite{DKPS97}, but a quantitative understanding of the
strong coupling regime is lacking. Even with $N=2$, the matrix model
will be a highly non-trivial many body system. One fact which suggests
this is that Monte Carlo simulations of supersymmetric matrix quantum
mechanics~\cite{Han10, Hanada} with $N$ as small as 2 or 3 yield 
results consistent 
with the predictions of gauge/gravity correspondence~\cite{SeYo00,
Sekino} which is supposed to be valid in the large $N$ limit.  

One cannot really define scattering amplitudes in our bosonic model,
since there are no weakly-interacting asymptotic states without
supersymmetric cancellations. However, our model may capture the essence
of the strong coupling regime where many excited states contribute to
the dynamics, since those states will not be sensitive to the detailed
features of the model such as supersymmetry. Also, 
the gauge symmetry has an
important consequence: D-branes are repelled from each other at short
distances because of the centrifugal force due to angular momenta in the
gauge space, as will become clear in our approach. 

Even though the $d=2$, $N=2$ bosonic model (see eq.~\eqref{eq:4H} below) looks simple, it is not integrable (in the classical sense), and features rich dynamics. This model has been studied before under the name of Yang-Mills quantum mechanics \cite{Sa83, Sa84, Fu87, JS89}, and is considered to be a prototypical model for quantum chaos. Despite intensive numerical simulation, we still lack a complete understanding even in the classical realm. The chaotic nature of Yang-Mills theory may provide insight into the problems such as thermalisation effects in hadronic processes \cite{KMOSTY10} and confinement in Yang-Mills theory \cite{Ole82}.

In this work we take a fresh approach to the problem, by first
presenting a new equivalent formulation, taking full advantage of the
gauge symmetry and exhibiting an aspect of locality in the model. It
turns out that in this reformulation --- which is is amenable to
analytical study and precise numerical analysis --- questions of
equilibration~\cite{Linden_etal09, CramerEisert08,
InteractingThermalisation} in non-equilibrium can be precisely
posed. Our considerations complement other studies, such as classical
analysis of the thermalisation in matrix
models~\cite{BerensteinClassical2, BerensteinClassical}, as well as the
bounds on scrambling time based on the locality of the models and
Lieb-Robinson bounds~\cite{FastScrambling}. 
See also recent work~\cite{Sahakian1, Sahakian2, Sahakian3, 
Mandal, Kabat} for different
approaches to thermalisation in matrix models.


\section{Matrix models and relations to other models}

We will study bosonic matrix models defined by the Hamiltonian of $0+1$ dimensional Yang-Mills theory
\begin{equation}
	H = {\mathrm{Tr}} \biggl[{\frac{1}{2} \sum_{i=1}^d \mathfrak{P}^{(i)} \mathfrak{P}^{(i)} -
	\frac{1}{4}\sum_{i,j=1}^d [\mathfrak{X}^{(i)}, \mathfrak{X}^{(j)}]^2}\biggr],
\label{eq:Hmatmod}
\end{equation}
where $\mathfrak{X}^{(i)}= \sum_a x^{(i)}_a t_a$, with the $N \times N$
traceless Hermitian generators $t_a$ of $SU(N)$, and position operators
$x^{(i)}_a$. Similarly we write $\mathfrak{P}^{(i)}= \sum_a p^{(i)}_a
t^*_a$ where $p^{(i)}_a$ are the conjugate momenta. The gauge field is
not dynamic and has been set to zero by a gauge transformation; its
equation of motion imposes the constraint that states under
consideration are required to be singlets of $SU(N)$. In this work we
will primarily treat the $d=2$, $N=2$ case\footnote{The $d=2$ model is
given by a dimensional reduction of pure Yang-Mills theory in $2+1$
dimensions (see ref. \cite{Leigh} for a recent study). The supersymmetric
version of the $d=3$, $N=2$ model has been studied in ref. \cite{Wosiek}
using a method similar to ours.}.
An important theorem which provides a starting point for
our analysis is that the
Hamiltonian \eqref{eq:Hmatmod} with the gauge constraint can be
recast into a model which has a local structure of interactions.
\begin{theorem}[Equivalent model]
The Hamiltonian \eqref{eq:Hmatmod} for $N=2$ and $d=2$, with the constraint that the states are SU(2) singlets, is equivalent with
\begin{equation}
\label{eq:easyHam}
\begin{split}
	4 H =& \left(P^2 \otimes \one +\one \otimes P^2 \right) \otimes \one \\
	&+
	\left(
	X^{-2} \otimes \one+ \one \otimes X^{-2}
	\right) \otimes \sum_{\ell} {\ell(\ell+1) |\ell\rangle\langle \ell|} \\
	&+
	X^2 \otimes X^2
	 \otimes \sum_{\ell,\ell'} A_{\ell,\ell'}\ketbra{\ell}{\ell'},
\end{split}
\end{equation}
acting on $\mathcal{H}= L^2( \mathbb{R}^+)\otimes L^2( \mathbb{R}^+)\otimes l^2$
where $A=A^T$ is banded, with
\begin{equation}
\label{eq:Amatrix}
\begin{split}
	A_{\ell, \ell} &= \frac 1 2-\frac{1}{8 \ell (\ell+1) - 6}\\
	A_{\ell, \ell+2} = A_{\ell+2, \ell} &= -\frac{(\ell+1)(\ell+2)}{\sqrt{2 \ell+1}(2 \ell + 3)\sqrt{2 \ell +5}},
\end{split}
\end{equation}
all other elements being zero.
\end{theorem}


\paragraph{Proof of equivalence.}
To prove the equivalence of the above models, we start from the explicit form of 2$\times$2 traceless Hermitian matrices, $\mathfrak{X}^{(i)} = \sum_{a=1}^{3} x^{(i)}_a \sigma_a/2$ with Pauli matrices $\sigma_a$. Expressing $\mathfrak{P}^{(i)}$ similarly, we have a set of pairs of canonical coordinates, $(x^{(i)}_a,p^{(i)}_a)$, whose Hamiltonian \eqref{eq:Hmatmod} is now written as
\begin{equation}
	4 H = (\mathbf{p}^{(1)})^2 + (\mathbf{p}^{(2)})^2 +
	 (\mathbf{x}^{(1)} \times \mathbf{x}^{(2)})^2. 
\label{eq:4H}
\end{equation}
In what follows, it is convenient to turn to the position representation in radial coordinates, in which the interaction term $(\mathbf{x}^{(1)} \times \mathbf{x}^{(2)})^2$ is identified with $r_1^2 r_2^2 \sin^2 (\theta_{1,2})$ where $\theta_{1,2}\in[0,\pi)$ is the angle between $1$ and $2$ and $r_1,r_2\in (0,\infty)$. Singlet states are given by linear combinations of 
\begin{equation*}
\psi^{(0)}_{\ell}(\theta_1, \phi_1; \theta_2, \phi_2) = (2 \ell +1)^{-1/2} \sum_{m=-\ell}^{\ell} (-1)^{\ell-m} Y_\ell^m(\theta_1, \phi_1) Y_\ell^{-m}(\theta_2, \phi_2),
\end{equation*}
with the spherical harmonics $Y_\ell^m:[0,\pi)\times [0,2\pi)\rightarrow \mathbb{C}$. Accordingly, the state vector in the position representation takes the form
\begin{equation*}
\psi(r_1, \theta_1, \phi_1; r_2, \theta_2, \phi_2) := \frac 1 {r_1 r_2} \sum_{\ell} \rho_{\ell}(r_1,r_2) \psi^{(0)}_{\ell}(\theta_1, \phi_1; \theta_2, \phi_2),
\end{equation*}
with radii-dependent functions $\rho_{\ell}:(0,\infty)\times(0,\infty)\rightarrow \mathbb{C}$.

The Hamiltonian in the radial position representation becomes
\begin{equation*}
-\partial_{r_1}^2 - \partial_{r_2}^2 - \frac{\Delta_{S_1}}{r_1^2} - \frac{\Delta_{S_2}}{r_2^2} + r_1^2 r_2^2 \sin^2 (\theta_{1,2}),
\end{equation*}
where $\Delta_S$ is the Laplace operator on the unit sphere, and $\theta_{1,2}$ is the angle on the great circle. The interaction term, which itself is invariant under $SU(2)$, can be rewritten using the identity
\begin{equation*}
\sin^2 (\theta_{1,2}) = \frac{8 \pi}{3} \left( \psi^{(0)}_{0}(\theta_1, \phi_1; \theta_2, \phi_2) - \frac 1 {5^{1/2}} \psi^{(0)}_{2}(\theta_1, \phi_1; \theta_2, \phi_2) \right).
\end{equation*}
Its action on the state vector can be found from the multiplication rules for spherical harmonics, $\psi^{(0)}_{2} \psi^{(0)}_{\ell} = \sum_{\ell'} c_{\ell,2,\ell'} \psi^{(0)}_{\ell'}$. The coefficient
\begin{equation*}
	c_{\ell,2,\ell'} = (-1)^{\ell - \ell'} \frac{5^{1/2}}{8 \pi}\left({(2 \ell' + 1)(2 \ell+1)}\right)^{1/2} \int_{-1}^{1} \mathrm{d}x P_{\ell}(x) P_2(x) P_{\ell'}(x),
\end{equation*}
is closely related to Wigner's 3J-symbol
\begin{equation*}
	\int_{-1}^{+1}\mathrm{d}x P_{\ell}(x) P_2(x) P_{\ell'}(x) =
	2 \left( \begin{array}{ccc}\ell & 2 & \ell' \\ 0&0&0 \end{array} \right)^2.
\end{equation*}
It is non-zero only when $\ell=\ell'$ or $\ell=\ell'\pm 2$. In the end, the Hamiltonian as an operator acting on $\rho$ takes the form of \eqref{eq:easyHam} with
\begin{equation*}
3A_{\ell, \ell'} = \left( 2 - c_{\ell,2,\ell} \right) \delta_{\ell,\ell'} - c_{\ell,2,\ell'} \delta_{\ell, \ell'+2} - c_{\ell,2,\ell'} \delta_{\ell, \ell'-2},
\end{equation*}
with the explicit form of $A_{\ell, \ell'}$ given by \eqref{eq:Amatrix}.\qed


\paragraph{Discussion of the model.}
Interestingly, the bandedness of $A$ introduces a notion of locality to the model not apparent in the original form eq.~\eqref{eq:Hmatmod}. There is a fast convergence $A_{\ell,\ell} \rightarrow 1/2$, $A_{\ell,\ell \pm 2} \rightarrow - 1/4$, also $\ell(\ell+1) \approx \ell^2$, for growing values of $\ell$. Hence, the part acting on $l^2$ reminds of the harmonic oscillator on a lattice~\cite{CGM86} where $A$ is an approximation of the Laplacian. The even and odd sublattices are decoupled, as there are no non-zero matrix elements coupling $\ell$ and $\ell \pm 1$. The dynamics on $L^2( \mathbb{R}^+)\otimes L^2( \mathbb{R}^+)$ determines joint effective ``mass'' and ``spring constants'' for the two indirectly coupled systems in $l^2$, providing indirect interaction between these systems.

The model is invariant under the spatial $SO(2)$ rotation. The generator in the representation of eq.~\eqref{eq:4H} is $\mathbf{x}^{(1)}\cdot \mathbf{p}^{(2)} -\mathbf{x}^{(2)}\cdot \mathbf{p}^{(1)}$, and we have an equivalent operator $Q$ in the representation of eq.~\eqref{eq:easyHam}
\begin{equation*}
\begin{split}
	Q =& (X \otimes P - P \otimes X)\otimes \sum_{\ell, \ell'}
	(Q_p)_{\ell, \ell'}|\ell\rangle\langle \ell'|\\
	&+ (X \otimes X^{-1} - X^{-1} \otimes X)\otimes \sum_{\ell, \ell'}
	(Q_x)_{\ell, \ell'}|\ell\rangle\langle \ell'|,
\end{split}
\end{equation*}
where $(Q_p)_{\ell,\ell+1} =(Q_p)_{\ell+1,\ell} = -\ell(4\ell^2-1)^{-1/2}$ and $(Q_x)_{\ell,\ell+1}=(Q_x)^*_{\ell+1,\ell} = -i \ell^2(4\ell^2-1)^{-1/2}$, all other elements being $0$. Note that because of the invariance of the model under the action of $Q$, we have $[Q,H]=0$, and the Hilbert space is divided into eigenspaces of $Q$. The spectral values of $Q$ are $q \in 2\mathbb{Z}$, the even integers. This angular momentum will become important in the analysis of the model.


\section{The spectrum}

\paragraph{Numerical analysis.}
The aim of our numerical analysis is the low energy regime of the model. We compute the spectrum of $H$ in the different eigenspaces of $Q$, as well as the spatial extension of the state, $\langle X^2 \otimes \one \otimes \one + \one \otimes X^2 \otimes \one\rangle^{1/2}$. The Hilbert space of this model is infinite dimensional, so we make use of a finite-dimensional approximation, which is the only approximation we make. We stress that without the presented reformulation, already the smallest non-trivial instance of the model --- $N=2$ and $d=2$ --- seems inaccessible even on super-computers, and it would be also difficult to implement the gauge constraint numerically. We restrict $L^2(\mathbb{R}^+)$ to the Hilbert space spanned by the first $h_0=107$ odd eigenfunctions of the harmonic oscillator, and we restrict the dimension of $l^2$ to some value $\ell_0=156$. The eigenvalues of $Q$ are $q \in 2 \mathbb{Z}$, the even integers. We enumerate the energies within each $Q$-eigenspace with a parameter $n \in \mathbb{N}^0$, hence we have states $\ket{q,n}$ with $Q\ket{q,n}=q\ket{q,n}$ and $H\ket{q,n}=E_{q,n}\ket{q,n}$. The quality of the approximation is double-checked by choosing smaller dimensions initially and then keeping only the the states whose eigenvalues remain almost unchanged when enlarging the dimension.


\paragraph{Results.}
The ground state energy is $E_{GS}=1.05535\ldots$

\begin{figure}[tb]\centering
\includegraphics[width=0.75\columnwidth]{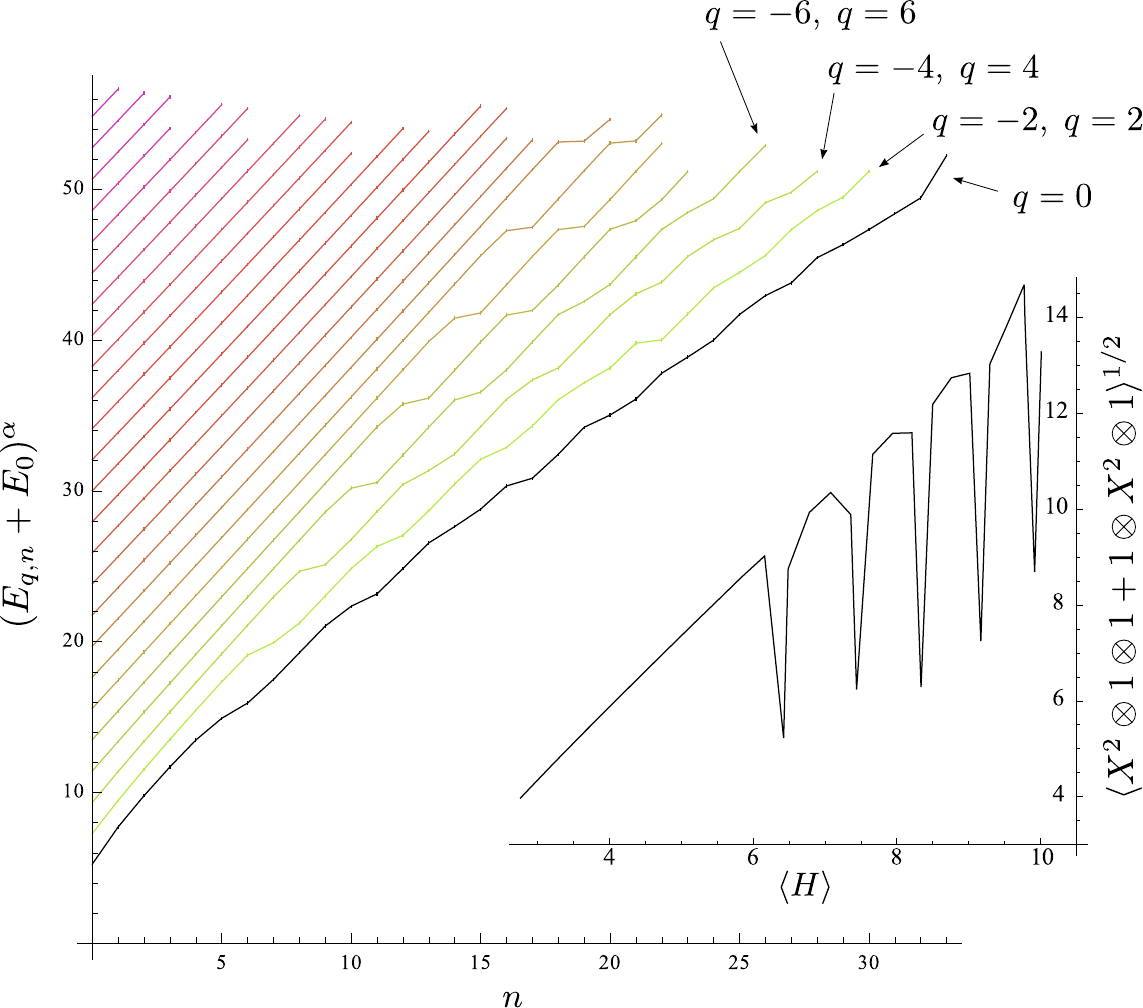}
\caption{(Large plot: energies)~$H\ket{q,n}=E_{q,n}\ket{q,n}$. Depicted is the value of $(E_{q,n} + E_0)^\alpha$ on the vertical axis with $n$ on the horizontal axis, using $\alpha \approx 1.62$. The black line belongs to $q=0$ and the others to increasing $|q|$, from green to red. There is affine growth of $(E_{q,n}+E_0)^\alpha$ over both $n$ and $|q|$ in certain regions. The affine growth over $n$ stops at certain points where we notice almost degenerate pairs of energies. The growth over $n$ is always sublinear.
(Small plot: sizes)~Depicted is the value of $\langle X^2 \otimes \one \otimes \one + \one \otimes X^2 \otimes \one \rangle^{1/2}$ for fixed $q=3$, as an example. This value is approximately an affine function of $\langle H \rangle$, with occasional collapses to much smaller values at the positions of the almost degenerate energies.}
\label{fig:qplot}
\end{figure}

The data suggest power laws both in $q$ and $n$. Figure~\ref{fig:qplot}
shows, for fixed $q$, an affine function $(E_{q,n} + E_0)^\alpha$ up to
certain values of $n$, where we notice almost degenerate pairs. After
that, linear growth continues. Low values of $|q|$, especially $q=0$,
are dominated by the degeneracies and hence the linear growth is mostly
hidden, but the onset of degeneracies comes later for larger values of
$|q|$. One can fit the data for fixed $n$ as well, the dependency seems
correct for a range of exponents $\alpha \approx 1.5 \ldots 2.3$,
provided the other constants are chosen appropriately. For $\alpha = 2$
this  relation is known as the linear Regge trajectory and corresponds
to the behaviour of a relativistic string~\cite{GSW}, which has been
observed in the $2+1$ dimensional Yang-Mills
theory~\cite{PowerLaw}. There is a scaling argument coming from
semi-classical analysis which suggests the dependence $E_{q,0} \sim
|q|^{2/3}$~\cite{BerensteinScaling}.
The best fit we obtain when leaving all constants subject to
variation yields $\alpha \approx 1.62(2)$, $E_0 \approx 1.6(1)$,
but we cannot draw a definite conclusion about the value of $\alpha$
from our analysis alone. Note
that the growth of $(E_{q,n}+E_0)^\alpha$ with $n$ stays always
sublinear for all values of $q$. We also consider the spatial extension
of the states, $\langle X^2 \otimes \one \otimes \one + \one \otimes X^2
\otimes \one\rangle^{1/2}$. The extensions within each eigenspace of $Q$
are affine functions of $E$, except that around the degenerate pairs
states have a much smaller size. 

\begin{figure}[tb]\centering
\includegraphics[width=0.75\columnwidth]{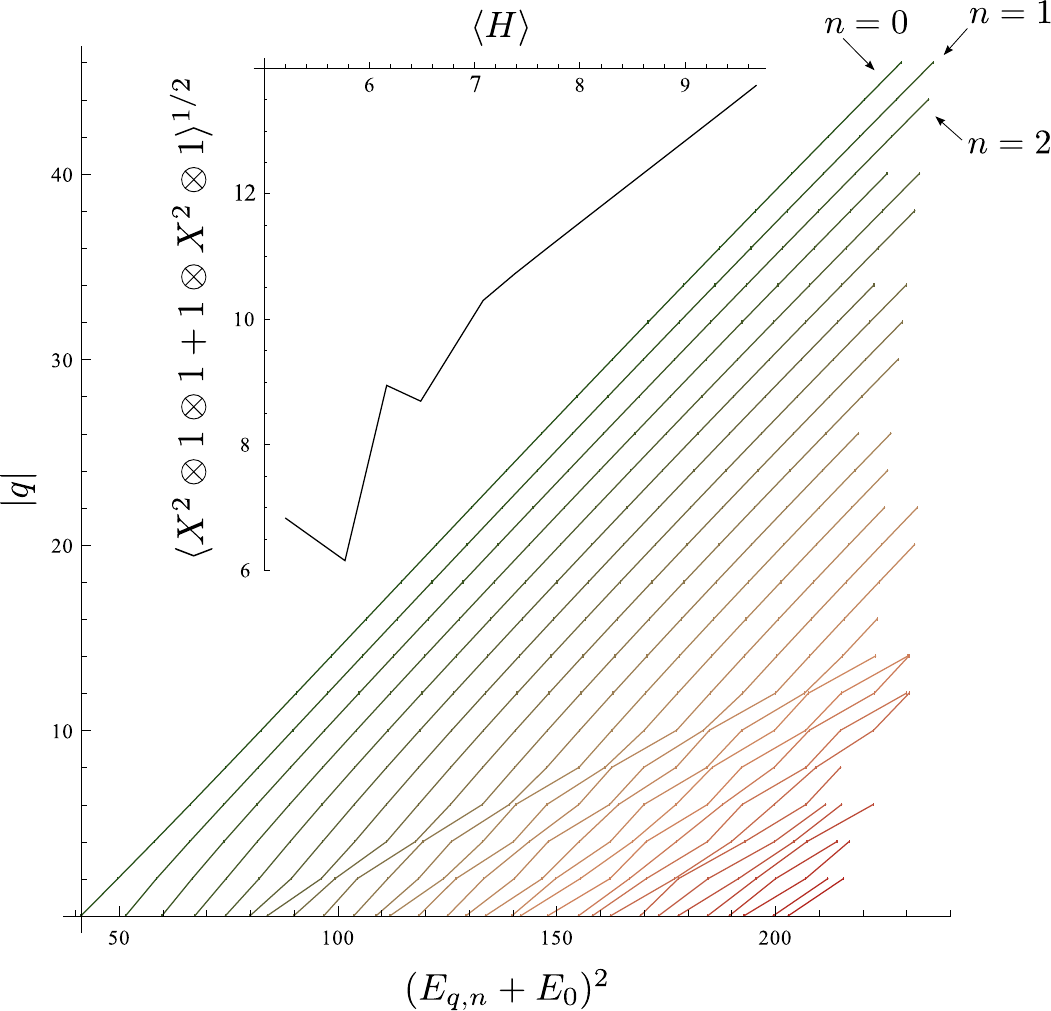}
\caption{Chew-Frautschi plot and size plot (large plot: energies). Depicted is the value of $(E_{q,n} + E_0)^2$, similarly to Figure~\ref{fig:qplot}, but fitting the data while fixing $\alpha=2$. In this plot, we show graphs belonging to constant values of $n$, while varying over values of the angular momentum $q$. An asymptotic affine dependence between $(E_{q,n} + E_0)^2$ and $|q|$ is clearly seen in the large energy / large angular momentum regime.
(Small plot: sizes)~Depicted is the value of $\langle X^2 \otimes \one \otimes \one + \one \otimes X^2 \otimes \one \rangle^{1/2}$ for fixed $n=11$, as an example, over the full range of $q$. This graph follows, asymptotically for large $|q|$ and $n$, an affine dependence as well, with a low energy region where the size and energy are smaller than a linear extrapolation of the asymptotics implies. Is resembles the behaviour of an initially strongly bound system which is less tightly bound for high energies.}
\label{fig:CFplot}
\end{figure}

In Figure~\ref{fig:CFplot} we show a Chew-Frautschi plot of the model. An asymptotic affine dependence between $(E_{q,n} + E_0)^2$ and $|q|$ is clearly seen in the large energy / large angular momentum regime.


\section{Equilibration in matrix models} 

Equipped with these insights, we now turn to questions of equilibration. In the quantum setting, `equilibration' refers to the situation that for most times, expectation values take values as if the system was in the time-averaged state
\begin{equation}
\omega:= 
\lim_{T\to \infty}
\frac{1}{T}\int_0^T \mathrm{d}t e^{-itH} \rho_0 e^{itH},\qquad \rho_0:= |\psi(0)\rangle\langle\psi(0)|.
\label{omega}
\end{equation}
Here, we consider the sectors $H_q=\sum_n E_{q,n} \ketbra{q,n}{q,n}$ of
the Hamiltonian $H$ for a given value of $q$. Assuming the 
non-degeneracy of energy levels, the time-averaged state \eqref{omega}
is diagonal in the energy basis,
\begin{equation}
 \omega=\sum_{n}|\langle \psi(0)|q,n\rangle|^2
|q,n\rangle\langle q,n|.
\end{equation}
A measure of the extent to which expectation values stay close to 
those of the time average is
the effective dimension $d_{\rm eff}^{-1} := {\rm Tr}\, \omega^2
=\sum_n |\langle
\psi(0)|q,n\rangle|^4$.
In fact, if we divide the system into a $d_S$-dimensional subsystem $S$ 
and the rest (which is often referred to as the heat bath) $B$,
the reduced density matrix $\rho_S(t)={\rm Tr}_B[\rho(t)]$ 
is close to the time average $\omega_S={\rm Tr}_B[\omega]$ if
$d_{\rm eff}$ is large~\cite{Linden_etal09}, 
\begin{equation}
	\mathbb{E}\|\rho_S(t)-\omega_S\|_1 \leq d_S / d_{\rm eff}^{1/2},
\label{rhoS}
\end{equation}
where $\mathbb{E}$ denotes the time average, and the distance in the 
state space is measured with respect to the trace-norm, 
$\|M\|_1={\rm Tr}\sqrt{M^\dagger M}$.

We can immediately get an estimate for the
effective dimension of a random state in a micro-canonical energy
window, slightly improving on ref.~\cite{Linden_etal09}. Denote with
$d_\Delta$ the number of spectral values of $H_q$ contained in an
interval $[E,E + \Delta]$ for a given fixed value of $q$, and let $I$
denote the corresponding index set $k \in I \Leftrightarrow E_{q,k} \in
[E,E+\Delta]$ in this eigenspace of $Q$. 

\begin{observation}[Effective dimensions] A Haar random state vector $\ket{\psi(0)}$ in the micro-canonical subspace spanned by $\{\ket{{q,n}}: n \in I\}$ satisfies
	$\mathbb{E}(d_{\rm eff}) \geq (1+ d_{\Delta})/2$.
\label{obs:effdim}
\end{observation}
Using the Weingarten function calculus~\cite{Collins1, Collins2}
for computing moments of entries
of Haar-random unitaries $U$, 
and making use of the fact that each entry
is identically distributed, one finds 
\begin{multline*}
\mathbb{E} \left( d_{\rm eff}\right) = \mathbb{E} \left[
 \frac{1}{\sum_{n \in I}|\langle q,n| U| q,n_{1}\rangle|^4} \right] \\ 
	\geq \frac{1}{\mathbb{E}\left [
	\sum_{n \in I}|\langle q,n | U| q,n_{1}\rangle|^4
	\right]}
	= \frac{1}{ d_\Delta \mathbb{E}\left [
	|\langle q,n_{1} | U| q,n_{1} \rangle|^4
	\right]} = \frac 1 {d_\Delta} \binom{d_\Delta+1}{2}= \frac{d_\Delta+1}{2}.
\end{multline*}
where $|q, n_{1}\rangle$ is one of the state vectors in the microcanonical
subspace.
\qed

One can also study the equilibration of observables.
Given the assumption of the non-degeneracy of energy gaps, 
the following relation~\cite{Short1} holds for any operator $A$,
\begin{equation}
\mathbb{E}\left[
|{\rm Tr}[\rho(t)A]- {\rm Tr}[\omega A]|^2\right]
\leq {\|A\|^2\over d_{\rm eff}},
\label{rhoA}
\end{equation}
where $\mathbb{E}$ denotes the time average, 
and $\|A\|$ is the operator norm, defined as the
largest singular value of the operator. 
If $d_{\rm eff}$ is large,
the outcome of a measurement represented by
the operator $A$ will be close to what we would get
in the time-averaged state $\omega$, most of the time.
In fact, the relation (\ref{rhoS}) can be derived from 
(\ref{rhoA})~\cite{Short1}, and it
has been shown that equilibration occurs in a finite 
time~\cite{Short2}. However,
having a large number $d_\Delta$ of energy levels of $H_q$ within an
energy window $[E,E+\Delta]$ implies that there are observables
that will take long to equilibrate. 

\begin{observation}[Equilibration times] Consider the initial state vector 
\begin{equation*}
|\psi(0)\rangle = d_\Delta^{-1/2}\sum_{n \in I} \ket{q,n},
\end{equation*}
which implies $d_{\rm eff}=d_\Delta$, and the observable
\begin{equation*}
O = \sum_{i,j \in I}(\delta_{i,j-1} + \delta_{i,j+1}) |q,i\rangle\langle q,j|.
\end{equation*}
Then the deviation from the infinite time average $\omega = \sum_{i \in
 I} \ketbra{q,i}{q,i}$, normalised by the operator norm  
of the observable $\|O\| \leq 2$, fulfills
\begin{equation*}
	\frac{|\Tr{O(\rho(t)-\omega)}|}{\|O\|}
	\geq 1 - \frac{1}{2 d_\Delta} \sum_{(i,i+1) \in I \times I} (E_{q,i+1} - E_{q,i})^2 t^2.
\end{equation*}
\label{obs:eqtimes}
\end{observation}

This follows immediately from the fact that the left hand side is bounded from below by
\begin{equation*}
	\sum_{i,j \in I} \frac{\delta_{i,j+1} + \delta_{i,j-1}}{2 d_\Delta} e^{-i t (E_{q,i} - E_{q,j})} = \frac 1 {d_\Delta} \sum_{(i,i+1) \in I \times I} \cos t (E_{q,i+1} - E_{q,i}),
\end{equation*}
which in turn is bounded from below by the right hand side.
\qed

We now apply Observations 2 and 3 to the matrix model
under consideration. 
We assume that $H_q$ is not degenerate. This is supported by the
numerical results, although step-like features in the spectrum exist,
which are very close to degeneracies. This does not, however, invalidate
the following argument unless a degeneracy is exact. Furthermore, there
is strong evidence presented above that within each eigenspace of $Q$ we
have sublinear growth of $(E_{q,n} + E_0)^{\alpha}$ with $n$, where
$\alpha=1.62$ in Figure~\ref{fig:qplot}. 
Hence for $\Delta \ll E$, we have $d_\Delta \geq \alpha \Delta (E+E_0)^{\alpha-1}/c$, where $c$ is the $n$-proportionality factor in $(E_{q,n}+E_0)^{\alpha} = a + b |q| + c n$, such that, with Observation~\ref{obs:effdim},
\begin{equation*}
	\mathbb{E}(d_{\rm eff}) \geq \frac{\alpha \Delta (E+E_0)^{\alpha - 1}}{2 c}+ \frac{1}{2}.
\end{equation*}
That is to say, for large initial energies, one expects a strong equilibration and expectation values often take the values close to the ones of the time average.

The sublinear growth of $(E_{q,n} + E_0)^{\alpha}$ allows to find an upper limit for the gaps in the spectrum. We find that for energies within one eigenspace of $Q$ and above $E$ we have $(E_{q,i+1} - E_{q,i}) \leq c / \alpha (E+E_0)^{\alpha-1}$, so we can estimate with Observation~\ref{obs:eqtimes} that
\begin{equation*}
	\frac{|\Tr{O(\rho(t)-\omega)}|}{\|O\|} \geq 1 - \frac{c^2 (d_\Delta-1)}{2 \alpha^2 d_\Delta} \frac {t^2}{(E+E_0)^{2(\alpha - 1)}}.
\end{equation*}
So for this specific initial condition, the system will be close to the infinite time average in expectation, but we can make the equilibration time scale as large as we want by shifting the energy up. These statements can only be concluded from the microscopic Hamiltonian once the spectrum as above has been identified, as facilitated by our new normal form.


\section{Conclusions}

\paragraph{Paths towards generalisations.}
To establish a relation between matrix models and black hole physics, studies of $d \ge 3$ will be important, since gravity in lower dimensions is special. The case of higher $N$ and $d$ is under study. Hamiltonian \eqref{eq:Hmatmod} with general $N$ and $d$ can be mapped to a model which involves $d$ particles in a $N^2-1$ dimensional space of the adjoint representation of $SU(N)$. The analysis will be more complicated than in $N=2$, $d=2$, since one needs more than one quantum number to specify a singlet wave function. However, the local structure of interactions persists, which has played an essential role in the analytic and numerical treatment in this paper, since the interaction terms in the Hamiltonian can change the $SU(N)$ quantum numbers only by a certain amount.


\paragraph{Summary.} We have constructed a model that is equivalent to the $N=2$, $d=2$ bosonic matrix model. We studied its spectrum as well as its equilibration properties. We found an affine dependence of $(E_{q,n} + E_0)^\alpha$, with plausible $\alpha \approx 1.5 \ldots 2.3$, on $|q|$ and $n$ in certain regions. The dependence on $|q|$ for the lowest $n$ (the leading Regge trajectory) is well-known to be the behaviour of a string if $\alpha=2$, and so is the spatial extension, which we found to be proportional to the energy (except for certain states). It is remarkable that this kind of non-perturbative behaviour, which is usually found with extensive Monte Carlo simulations in lattice gauge theory, is found by straightforward diagonalisation in our approach.

Although the $1/N$ expansion of any gauge theory can be represented as strings~\cite{tHooft}, it is far from obvious what kind of string theory our model with $N=2$ should correspond to. In fact, the power law dependence on $n$ without any degeneracy (up to some $n$, for fixed $q$) resembles a system of one or two oscillators, such as 
ref.\ \cite{Yukawa:1950eq}, rather than strings, which have an infinite number of modes. The states at larger $n$ that have small spatial extension may represent composite states.

Regarding the question of equilibration, we applied rigorous
mathematical results, which, complemented by the numerical analysis
regarding the spectrum, allowed us to estimate how the effective 
dimension grows as the energy of a micro-canonical subspace increases.
Our result shows that for high energies, states will for
the overwhelming proportion of times look like their long time average.
We also pointed out that observables and state preparations exist
where the equilibration takes arbitrarily long. 
In future investigations, these observables might be used to exhibit
aspects of the complexity of states  
as it builds up over time. Linked
(but not limited) to the notion of classical computational 
complexity, this means essentially the number of computation steps or
other resources used to construct a state from a 
simpler one (see, e.g., ref. \cite{Nielsen1, Nielsen2, Briegel}).
In fact, there is a recent proposal about a connection between 
the complexity of a quantum state and the 
interior geometry of a black hole, namely the volume of the 
Einstein-Rosen 
bridge~\cite{SusskindComplexity1, SusskindComplexity2}.
Complementing this development, there has been a recent activity on
the relationship between equilibration properties of Hamiltonians and the 
complexity of their eigenvectors, quantified in terms of the
length of a quantum circuit that is required to prepare a given quantum state
 \cite{Complexity}, closely related to notions of quantum Kolmogorov
 complexities \cite{Ge,Kolmogorov}.
 We suspect 
operators of the type we studied may help in describing the 
physics in the black hole interior.

We regard our analysis to be a first step toward understanding 
the behaviour of BFSS matrix theory,
treated as a small-dimensional, but fully
quantum mechanical model. The local structure of 
interaction in the space of gauge-invariant states has been
an essential tool in our analysis, but its physical consequence
is yet to be understood. We hope this locality sheds some light
on the unsolved problem of describing local spacetime physics
in quantum gravity.

\paragraph{Acknowledgements.} We thank S.\ Shenker for discussions and
J.\ Plefka, D.\ Berenstein, T. Yoneya, J. Nishimura and L. Susskind
for helpful
comments. This work has been supported by the EU (SIQS, Q-Essence,
RAQUEL), the FQXi, the ERC (TAQ), and the JSPS. This work was supported
in part by the National Science Foundation under Grant No.\ PHYS-1066293
and the hospitality of the Aspen Center for Physics. 
\paragraph{Note added.} The recent ref.\ \cite{Short} also addresses 
quantum processes slowly equilibrating, even though in a different context and providing different bounds.


\appendix


\section{Basics of gauge theory and BFSS matrix theory}

\subsection{Gauge theory}
Matrix quantum mechanics studied in this work is obtained from pure Yang-Mills theory in $d+1$ dimensions by a dimensional reduction (\ie by assuming that the fields depend only on time). The Lagrangian of the $d+1$ dimensional theory is given by
\begin{equation*}
	L_{d+1}=-\frac 1 4 \Tr{F_{\mu,\nu}F^{\mu,\nu}}
\end{equation*}
where the repeated indices $\mu, \nu=0,1,\ldots, d$ are summed over. The field strength is
\begin{equation*}
	 F_{\mu,\nu}=\partial_{\mu}A_{\nu}
	 -\partial_{\nu}A_{\mu}
	+i [A_{\mu}, A_{\nu}].
\end{equation*}
Gauge field components $A_{\mu}$ are associated with the Lie algebra of the gauge group; for $SU(N)$, they are represented as traceless Hermitian $N\times N$ matrices. The gauge transformation by an $SU(N)$ element $U$ is given by
\begin{equation*}
	 A_{\mu}\mapsto i (\partial_{\mu} U)U^{-1} +U A_{\mu} U^{-1},
	\quad F_{\mu,\nu}\mapsto U F_{\mu,\nu} U^{-1},
\end{equation*}
and the Lagrangian is invariant under this transformation.
The Lagrangian of matrix quantum mechanics (Yang-Mills theory in $0+1$ dimensions) is obtained by setting the spatial derivatives to zero. Writing $A_{0}=A$ and $A_{i}=\mathfrak{X}^{(i)}$, the field strengths become
\begin{equation*}
	F_{0i}=D_{0} \mathfrak{X}^{(i)}=\dot{\mathfrak{X}}^{(i)}+i [A, \mathfrak{X}^{(i)}],
	\quad F_{ij}=i[\mathfrak{X}^{(i)}, \mathfrak{X}^{(j)}],
\end{equation*}
and we obtain
\begin{equation*}
	L = \Tr{\frac 1 2\left(\dot{\mathfrak{X}}^{(i)}
	+i[A, \mathfrak{X}^{(i)}]\right)^2
	+ \frac 1 4 [\mathfrak{X}^{(i)}, \mathfrak{X}^{(j)}]^2},
\end{equation*}
where the repeated indices $i, j=1, \ldots, d$ are summed over. The infinitesimal form of a gauge transformation for $U=e^{-i\Lambda}$ is
\begin{equation}
	\delta A=\dot{\Lambda} +i[A, \Lambda],
	\quad \delta \mathfrak{X}^{(i)}=i[\mathfrak{X}^{(i)}, \Lambda].
\label{eq:gaugetr}
\end{equation}
In $0+1$ dimension, the gauge field $A$ is not a dynamical field, since it has no kinetic term. We can set $A$ to zero by a gauge transformation; $A$ only plays the role of Lagrange multiplier, which imposes a constraint (equation of motion w.r.t.\ $A$) on the system. We will perform the canonical quantisation in the $A=0$ gauge. Consider each real number of the matrix elements as a dynamical variable. The trace part represents the center of mass. The corresponding momentum is conserved, and its dynamics is decoupled from the rest. So we will concentrate on the study of the relative motion.

For the explicit analysis, it is convenient to expand the fields in the basis $t_a$,
\begin{equation*}
	\mathfrak{X}^{(i)}=\sum_{a=1}^{N^2-1}x_{a}^{(i)}t_a,
	\quad A=\sum_{a=1}^{N^2-1}a_{a}t_a
\end{equation*}
where $t_{a}$ are traceless Hermitian matrices which satisfy
\begin{equation*}
	\Tr{t_{a}t_{b}}=\frac 1 2 \delta_{a,b},
	\quad [t_a, t_b]=i\sum_{c}f_{a,b,c}t_c,
\end{equation*}
and where $f_{a,b,c}$ is the structure constant of $SU(N)$. For the $SU(2)$ gauge group, studied in the main text, we have $t_a=\sigma_{a}/2$ and $f_{a,b,c}=\epsilon_{a,b,c}$. Gauge transformation \eqref{eq:gaugetr} for $x_{a}^{(i)}$ with $\Lambda=\sum_{a} \lambda_{a}t_{a}$ is
\begin{equation}
	\delta x^{(i)}_{a}= \sum_{b,c} f_{a,b,c}\lambda_{b}x^{(i)}_{c} .
\label{eq:xgauge}
\end{equation}
The conjugate momenta for $x_{a}^{(i)}$ are
\begin{equation*}
	p_{a}^{(i)}=\frac{\partial L}{\partial\dot{x}_{a}^{(i)}}
	=\frac{\partial}{\partial\dot{x}_{a}^{(i)}}\sum_{a}
	\frac{(\dot{x}_{a}^{(i)})^2}{4}=\frac 1 2 \dot{x}_{a}^{(i)},
\end{equation*}
and the Hamiltonian is
\begin{equation*}
	H=\sum_{a}(p_{a}^{(i)})^{2}
	+\frac 1 8 \sum_{a,b,c,d,e}
	f_{a,b,c}f_{a,d,e}x^{(i)}_{b}x^{(j)}_{c}x^{(i)}_{d}x^{(j)}_{e}
\end{equation*}
For $SU(2)$, by using $\sum_{a=1}^{3} \epsilon_{a,b,c}\epsilon_{a,d,e} =\delta_{b,d}\delta_{c,e} -\delta_{b,e}\delta_{c,d}$, and redefining $x_{a}^{(i)}\mapsto 2^{1/3}x_{a}^{(i)}$, $p_{a}^{(i)}\mapsto 2^{-1/3}p_{a}^{(i)}$ and $H\mapsto 2^{-4/3}H$ we get eq.\ \eqref{eq:4H}.
The equation of motion for $A$ (obtained by varying the Lagrangian by $a_{a}$ and setting $a_{a}=0$ afterwards) is
\begin{equation}
	0= \frac{\partial L}{\partial a_{a}}
	=\sum_{b,c}f_{a,b,c}p_{b}^{(i)}x_{c}^{(i)} \equiv V_a.
\label{eq:aeom}
\end{equation}
Noether's theorem tells us that the $V_a$ are the generators of $SU(N)$ transformations. We impose constraint \eqref{eq:aeom} by requiring that the physical state vectors $\ket{\psi}$ are annihilated by the generators $V_{a}$,
\begin{equation*}
	V_{a}\ket{\psi}=0.
\end{equation*}
In other words, $|\psi\rangle$ are singlets of $SU(N)$.

\subsection{BFSS matrix theory}
Banks-Fischler-Shenker-Susskind (BFSS) matrix theory is given by the dimensional reduction of conventional $U(N)$ Yang-Mills theory with $d=9$ as described above, supplemented by fermionic degrees of freedom, which make the theory supersymmetric. The bosonic degrees of freedom are $N\times N$ Hermitian matrices $\mathfrak{X}^{(i)}$, $i=1,\ldots, 9$. Their supersymetric partners are fermionic $N\times N$ matrices $\Theta_\alpha=\sum_a \theta_{\alpha,a} t_a$, where $\theta_{\alpha, a}$ ($\alpha=1, \ldots, 16$) are are Majorana-Weyl spinors in $9+1$ dimensions which have $16$ real components. This matrix quantum mechanics with the maximal amount of supersymmetry has been proposed as a formulation of M-theory, the strong coupling limit of type IIA string theory in (9+1) dimensions~\cite{BFSS96}.

The requirement of $SO(9)$ symmetry and maximal supersymmetry 
determines the action uniquely\footnote{If one breaks the SO(9) 
symmetry to $SO(3)\times SO(6)$, a maximally supersymmetric action 
with mass term exists, which is called the 
Berenstein-Maldacena-Nastase (BMN) matrix model~\cite{BMN}. 
Application of our method to this kind of mass-deformed
matrix models is an interesting subject for future investigations.}.
There is no mass term, and the 
interaction among
$\mathfrak{X}^{(i)}$ is quartic in the form of the trace of a commutator
squared. There are also cubic interactions involving two fermions and
one boson. The only parameter in the theory is the Yang-Mills coupling
$g_{YM}$ (which has a dimension of (length)$^{-3/2}$, being proportional
to the square root of the string coupling $g_s$). It appears in the
action as an overall factor, $1/g_{\rm YM}^2$. In our analysis, we have
set $g_{YM}=1$, since it can be scaled away by redefinitions of time and
the fields. 

In string theory, the model is interpreted as the description of a collection of $N$ so-called D0-branes at distances shorter than the string scale $\ell_s$, when their velocities are small. The diagonal elements of the matrices $\mathfrak{X}^{(i)}$ denote the position of the D0-branes, and their off-diagonal elements are interpreted as the lowest modes of open strings stretched between the D0-branes. The $U(1)$ part ($\mathfrak{X}^{(i)}$ being proportional to the unit matrix) corresponds to the center of mass motion, and is decoupled from the rest. Thus one can concentrate to the case of traceless matrices with $SU(N)$ gauge symmetry.

D-branes have features that are unfamiliar from ordinary systems due to the presence of the off-diagonal elements. First, the spatial coordinates are non-commutative. Second, there is $U(N)$ gauge symmetry, which means that the configurations that are related by $U(N)$ transformation are to be identified. This is in a sense a generalisation of fermionic and bosonic statistics. The consequences of these features are not fully understood.

Some evidence exists that the propose matrix quantum mechanics is a formulation of M-theory. M-theory is defined in a $(10+1)$ dimensional spacetime, where one of the spatial directions is compactified into a circle. The radius of compactification is $g_s \ell_s$, so in the strong coupling limit, the circle decompactifies. A D0-brane is considered to be a Kaluza-Klein mode which has one unit of momentum in the compactified direction. The BFSS conjecture is that the $N \to \infty$ limit with $g_s$ fixed describes M-theory in the infinite momentum frame (light-cone frame). In the light-cone frame the motion becomes non-relativistic, which gives a justification for the use of non-relativistic matrix quantum mechanics for its description. Another indication which suggest the connection between this theory and M-theory is that matrix quantum mechanics has an interpretation as a regularisation of super-membranes, which are believed to be important degrees of freedom in M-theory and are known to exist in $(10+1)$ dimensions. 

Some progress has been made which supports the interpretation of matrix theory as a theory of gravity. Many questions remain unanswered though, mainly because the model is very difficult to solve. Interactions between separated D0-branes have been studied by computing quantum effective actions around block diagonal configurations of matrices and shown to agree with the tree-level interactions of supergravity~\cite{OkYo99, Taylor}. In these studies, the distance $r$ between D0-branes introduces a mass scale, which allows one to compute the effective potential (or the phase shift in the Born-Oppenheimer approach) as a perturbative expansion in powers of $1/r$.
 
Supersymmetry plays a crucial role in the agreement of matrix and gravity calculations. Open string modes stretched between D0-branes have mass proportional to $r$, which gives rise to linearly increasing effective potentials when integrated out. Due to non-trivial cancellations of bosonic and fermionic fluctuations one can obtain a behaviour similar to massless gravitons, $1/r^{d-2}$ in $d$ spatial dimensions. (When D0-branes are at rest with respect to each other, exact cancellation occurs and there is no force between them.) In the bosonic model that we study, the effective potential grows linearly, and all the states will be bound states.

For an interpretation as M-theory it is needed that the maximally supersymmetric matrix models with any $N$ has a normalisable ground state with exactly zero energy. This has been shown for $N=2$ in 
ref. \cite{SeSt98} by computing the Witten index. There are also attempts to construct the ground state wave functions. See ref. \cite{Yin10} for a recent work. It is known that the theory has a continuous spectrum~\cite{DLN89}. It has been considered to be a problem before the appearance of the BFSS conjecture since it suggests super-membranes are unstable, but now the continuous spectrum is regarded as a consequence of the theory describing a many-body system.

Black holes should appear in matrix theory in the form of localised excited states. Studies of the theory at finite temperature have been performed using Monte Carlo methods. Internal energy and the spatial extent of the bound state have been computed and shown to be consistent with black hole thermodynamics~\cite{Nis98, Nis2}. Monte Carlo studies have been applied also to correlation functions in the zero-temperature limit~\cite{Han10, Hanada}, and the results agree in high precision with the prediction of the gauge/gravity correspondence. Interestingly, the results with $N$ as small as $2$ or $3$ agree with gravity results, too, which are, however, supposed to be valid mainly in the large $N$ limit.

Since we have strong evidence for the validity of matrix theory as a description of black holes, an important next step would be to describe black holes using a pure state and follow their dynamical evolution. The most obvious question about black holes concerns the microscopic origin of its entropy. Another question is the derivation of fast scrambling~\cite{HaPr07,Susskind} (thermalisation in a time logarithmic in the number of degrees of freedom, which is faster than in ordinary local field theory) from a fundamental theory. It is expected that matrix models realise fast scrambling because of the non-locality of interaction in the space of matrix elements, but no analysis of quantum dynamics has been made so far. (See ref. \cite{BerensteinClassical} for an interesting study of thermalisation in the classical limit.)

\bibliographystyle{JHEP}
\bibliography{MatrixModelsJHEPv2}

\end{document}